\documentclass[usenatbib]{mn2e}

\usepackage{bm}        
\usepackage[english]{babel}
\usepackage{amsfonts,amsmath,amssymb,mathrsfs,times,graphicx}

\newcommand{\eq}{\begin{equation}}
\newcommand{\eeq}{\end{equation}}

\title[Relativistic dynamical friction in a collisional fluid]{Relativistic dynamical friction in a collisional fluid}
\author[E.\ Barausse]{E.\ Barausse\thanks{E-mail: barausse@sissa.it}\\
SISSA, International School for Advanced Studies and INFN, Via Beirut 4, 34014 Trieste, Italy
}
\begin{document}

\date{}

\maketitle

\label{firstpage}

\begin{abstract}
The dynamical friction force experienced by a body moving at relativistic speed in a  
gaseous medium is examined. This force, which arises due to the gravitational interaction of the body with its own gravitationally-induced wake, 
is calculated for straight-line and circular motion, generalizing previous results by several authors. 
Possible applications to the study of extreme mass-ratio inspirals around  strongly-accreting supermassive black holes are suggested.
\end{abstract}


\section{\label{sec:intro}Introduction}

The  mechanism of dynamical friction (DF), which arises because of the gravitational interaction between a massive perturber moving in a medium and its own gravitationally-induced wake, 
was first  studied in collisionless systems by \citet{chandra}, and has had widespread applications in astrophysics [\textit{e.g.,} stars moving in clusters or galaxies, globular clusters in galaxies, galaxies in galaxy clusters, etc.: see \citet{BT}  section 7.1, and references therein]. In particular, the  Newtonian dynamical friction drag acting on a perturber of 
gravitational mass $M$ moving with velocity $\boldsymbol{v}_M$ in a collisionless system of ``particles'' with gravitational mass $m$ and isotropic velocity distribution $f(v_m)\equiv dN/(d^3{x}d^3{v}_m)$ is given \citep{chandra} by  
\begin{equation}\label{eq:chandra}
\boldsymbol{F}_{\rm DF}=-16\pi^2G^2M(M+m)\frac{m\int_0^{v_M}f(v_m)v_m^2dv_m}{v_M^3}\boldsymbol{v}_M\ln\Lambda
\end{equation}
where $\Lambda\approx b_{\max} v_{\rm typ}^2/[G(M+m)]$, $b_{\max}$ and $v_{\rm typ}$ being respectively the maximum impact parameter and the typical velocity of the particles with respect to the perturber [see also \citet{BT}  section 7.1 for a derivation]. The intuitive reason for the presence of this drag is the fact
that the particles are attracted by the perturber, which in the meantime moves: the particles therefore build up
a slight density enhancement behind it (the wake). It is the gravitational attraction of the wake that pulls the perturber back.
Note that in the case of a perturber moving through a collisionless fluid, 
dynamical friction is essentially the only drag force acting on the perturber,
besides that due to gravitational-wave emission 
[which is usually subdominant because it appears at 2.5 Post-Newtonian order:  see \citet{will_walker}, and also \citet{will_pati} and references therein] and that due to capture of 
particles by the perturber.

Equation \eqref{eq:chandra} was improved by including Post-Newtonian corrections in \citet{PNdf}, and was generalized
to relativistic velocities, although only in the weak scattering limit, in \citet{syer}.
The case of a  \textit{collimated} flow of collisionless particles of gravitational mass $m$ moving at 
relativistic speed and impacting on a perturber of gravitational mass $M$ 
was instead worked out by \citet{petrich}  [equation (B17)], who found that in the
rest frame of the perturber the 3-momentum change is given by
\eq\label{eq:non_coll}
\left(\frac{d\boldsymbol{p}}{dt}\right)_{\rm DF}=-\frac{4\pi m n_{\infty} G^2 M^2\gamma^2[1+(v/c)^2]^2}{v^2}\ln \left(\frac{b_{\max}}{b_{\min}}\right)
\frac{\boldsymbol{v}}{v}\,,
\eeq
where $n_{\infty}$ is the
number density of particles in the flow far away from the perturber and before deflection, 
$\boldsymbol{v}$ and $\gamma=[1-(v/c)^2]^{-1/2}$ are the  
velocity and the Lorentz factor of the  perturber  relative to the flow, and $b_{\min}$
is the size of the perturber or the capture impact parameter 
$b_{\min}\approx 2M(1+v^2)/v^2$ if this is a black hole.

Dynamical friction acts also in collisional fluids, together with the two other effects mentioned above for collisionless 
systems (\textit{i.e.,} gravitational-wave emission and accretion onto the perturber) and ordinary viscous forces, which are
\textit{not} present if the perturber is a black hole but instead act if the perturber is a star\footnote{The drag due to ordinary viscosity is 
given, for non-relativistic velocities and in the laminar regime, by Stokes' law: $\boldsymbol{F}_{\rm Stokes}=-6\pi \eta\, a \boldsymbol{v}$,
$a$ being the radius of the perturber and $\eta$ the viscosity coefficient.
For instance, in a thin accretion disc \citep{thin_disk} one has $\eta=\alpha \rho_0 c_s H$, 
where $\rho_0$ and $c_s$ are the rest-mass density and the sound 
velocity in the disc, $H$ is its height and $\alpha\sim 0.1 - 0.4$ \citep{alpha_value}. Note that this drag can
be calculated independently of the dynamical friction effects considered in this paper.}.
However, unlike in the
collisionless case, it presents different features depending on the Mach number of the perturber. The correct
behaviour in the supersonic case has long been recognized by several authors: the steady state Newtonian drag on a perturber $M$ moving on a straight-line
with velocity $\boldsymbol{v}$  relative to a homogeneous fluid with rest-mass density $\rho_0$ and sound speed $c_s=v/{\cal M}$ (${\cal M}>1$) 
was found by \citet{rephaeli_salpeter} and by \citet{wakes} to be
\eq\label{eq:supersonic_newt_drag}
\boldsymbol{F}_{\rm DF}=-\frac{4\pi G^2 M^2\rho_0}{v^2}\ln\left[\frac{b_{\max}}{b_{\min}}
\frac{{\cal M}}{({\cal M}^2-1)^{1/2}}\right]\frac{\boldsymbol{v}}{v}\,,
\eeq
where the maximum impact parameter $b_{\max}$ is the Jeans length (or the size of the medium if this is smaller than the Jeans length). 
Note that the dynamical friction drag given by equation 
\eqref{eq:supersonic_newt_drag} is comparable to the drag due to Bondi accretion onto the perturber: the
latter is in fact given by $\boldsymbol{F}_{\rm Bondi}=-\boldsymbol{v}\dot M$, where $\dot M={4\lambda\pi G^2 M^2\rho_0}/{(v^2+c_s^2)^{3/2}}$ 
\citep{bondi1,bondi2}, $\lambda$ being a parameter of order unity.

Equation \eqref{eq:supersonic_newt_drag}  was confirmed by \citet{ostriker} with a finite-time
analysis and was generalized to the relativistic case by \citet{petrich}, who found that equation \eqref{eq:non_coll} remains
valid also in the collisional supersonic case if the rest-mass density $m n_{\infty}$ is replaced by $p+\rho$, 
$p$ and $\rho$ being the pressure and energy-density of the fluid.
The physical reason for the presence of a non-zero drag in the supersonic case is the fact that sound waves can propagate only downwind,
inside the Mach cone, producing a non-symmetric pattern of density perturbations, which gives rise to a drag by gravitational interaction.

The subsonic case proved instead to be more elusive. Because sound waves can propagate both downwind and upwind, the drag is expected to be lower than
in the supersonic case. In particular, \citet{rephaeli_salpeter} in the Newtonian case and \citet{petrich} in the relativistic one argued that the
drag should be exactly zero for subsonic motion in a homogeneous fluid, because of the upwind-downwind symmetry of the 
stationary solution for the density perturbations excited 
by the perturber. However, although a zero drag can be a useful approximation in many cases, this result does not rigorously hold if one performs a 
finite-time analysis \citep{ostriker}. In fact, if the perturber is 
formed at $t=0$ and moves at non-relativistic subsonic speed on a straight-line in a homogeneous fluid, the 
density perturbations are given by the stationary solution found by \citet{rephaeli_salpeter} only inside a sphere 
of radius $c_s t$ centered on the initial position of the perturber, and
are instead exactly zero (because of causality) outside. The upwind-downwind symmetry of the stationary solution is therefore broken 
and the perturber experiences a finite drag, which reads \citep{ostriker}
\eq\label{eq:subsonic_newt_drag}
\boldsymbol{F}_{\rm DF}=-\frac{4\pi G^2 M^2\rho_0}{v^2}\left[\frac12\ln\left(\frac{1+{\cal M}}{1-{\cal M}}\right)-{\cal M}\right]\frac{\boldsymbol{v}}{v}
\eeq
as long as $(c_s+v)t$ is smaller than the size of the medium. This result was confirmed by numerical simulations \citep{simulations} and was
extended to the case of a perturber moving at non-relativistic speed on a circular orbit in a homogeneous medium by \citet{circ_drag}. 
In particular, \citet{circ_drag} found that in the subsonic case the perturber experiences a tangential drag, given roughly by 
equation \eqref{eq:subsonic_newt_drag}, and a drag in the radial direction (towards the center of the orbit),
whose contribution to the orbital decay is however subdominant with respect to the tangential drag. 
Similarly, in the supersonic case the tangential drag is roughly given 
by equation \eqref{eq:supersonic_newt_drag} with $b_{\max}$ equal to the orbital radius, while a radial drag is present but 
again remains subdominant with respect to the tangential one.

The purpose of this paper is to generalize to the relativistic case the finite-time drag found by \citet{ostriker} 
and  by \citet{circ_drag}. 
While a Newtonian treatment of dynamical friction is satisfactory in many astrophysical scenarios, 
relativistic expressions are needed in order to 
study the interaction of solar-mass compact objects or black holes 
with the gaseous matter (\textit{e.g.,} an accretion disc)  which 
could be present in the vicinity of a supermassive black hole (SMBH), where
orbital velocities close to that of light are reached. 
These systems, known as extreme mass-ratio inspirals [EMRIs; see \citet{emris} for a review], 
are expected to be among the most interesting sources of 
gravitational waves for the Laser Interferometer Space Antenna (LISA),
and considerable effort has been spent trying to understand
whether different   kinds of accretion disc, when present, can produce an observable 
signature in the emitted gravitational-wave signal. 
In a series of papers,  Karas, Subr 
and Vokrouhlicky considered the interaction between stellar satellites and
thin discs \citep{subr1,subr2,subr3,subr4}. \citet{subr4}, in particular, 
found that the effect of star-disc interaction on EMRIs
dominates over the effect of the loss of energy and angular momentum through gravitational waves 
in the case of thin discs, both
for non-equatorial orbits crossing the disc only twice per
revolution and for equatorial orbits embedded in the disc, unless
the orbiter is very compact (a neutron star or a black hole) or the
disc has a low density (\textit{e.g.,} in the region close to the
central SMBH if the flow becomes advection-dominated). These results
agree with those found by \citet{narayan}. He focused 
on Advection Dominated Accretions Flows [ADAFs \citep{ADAF}], which were then
believed to describe accretion onto ``normal'' galactic
nuclei\footnote{Nowadays, accretion onto ``normal'' galactic nuclei is
believed to be better described by Advection Dominated Inflow
Outflow Solutions (ADIOS) \citep{ADIOS}. However, this
is not expected to change significantly Narayan's results because ADIOS's, like ADAFs,
are expected to present very low densities in the vicinity the central SMBH.} 
(\textit{i.e.,} ones much dimmer
than Active Galactic Nuclei such as quasars, Seyfert galaxies, etc.).
He found that for compact objects and white dwarfs
the effect of the drag exerted by the accreting gas is negligible
compared with the loss of  energy and angular momentum through gravitational waves, 
whereas it is not negligible for main sequence and giant stars.
\citet{chakra1,chakra2} studied instead the orbital evolution of 
black hole satellites on circular equatorial orbits embedded in a
disc with a non-Keplerian distribution of angular momentum, and
found that the exchange of angular momentum between the disc and the
satellite can lead to important effects which have to be taken into
account when interpreting gravitational-wave signals from such
systems.

Although we do not expect our results to change significantly the picture outlined above for ADAFs or ADIOS's, whose density is too
low to make the effect of the hydrodynamic drag, and of dynamical friction in particular, comparable to the effects of
gravitational-wave emission even if one includes relativistic corrections, we think that our relativistic corrections could
play a more important role, under certain circumstances, for black holes or compact objects moving in higher density environments (Active Galactic Nuclei, quasars, Seyfert galaxies, etc.).
In a subsequent paper we will apply our results to the case of an accretion flow with a toroidal structure \citep{nextpaper}.

While our results rigorously apply only to a non self-gravitating fluid in either a flat background spacetime (in the case of 
straight-line motion) or the weak field region of a curved spacetime (in the case of circular motion), 
and additional work may be needed in order to evaluate the effect of a curved backround, we argue that such an approximation is 
suitable at least for a preliminary study of dynamical friction effects on EMRIs \citep{nextpaper}. Indeed, for many purposes 
a similar approximation is adequate to study gravitational-wave emission by EMRIs around a Kerr SMBH: 
the flat-spacetime quadrupole formula, combined with geodetic motion for the solar-mass satellite,
gives results which are in surprisingly good agreement with rigorously computed waveforms \citep{kludge}.

Our analysis closely follows that of \citet{ostriker} and \citet{circ_drag}, and we find that their results still hold for relativistic velocities
provided that the rest-mass density appearing in the Newtonian formulae is replaced by $p+\rho$ ($p$ and $\rho$ being the pressure and 
energy-density of the fluid), and a multiplicative factor is included: $\gamma^2[1+(v/c)^2]^2$ 
in the straight-line motion case and for the tangential component of the drag in the circular motion case; 
$\gamma^3[1+(v/c)^2]^2$ for the radial component of the drag in the circular motion case.

Throughout the rest of the paper we used units in
which $G = c = 1$. 

\section{\label{sec:equations}Equations and variables}

Let us consider a perturber with gravitational mass $M$, formed at $t=0$ and moving 
in a perfect fluid at rest at the initial position of the perturber\footnote{Note that this condition can always be satisfied
by performing a suitable boost.} and
having energy density $\rho$ and pressure $p$ there.
We write the metric as a Minkowski background plus some perturbations produced by the presence of the fluid and the perturber:
the general form of such a metric is known to be \citep{bardeen,kodama_sasaki,mukhanov,scott}
\begin{multline}
\label{eq:metric}
d\tilde{s}^2=\tilde{g}_{\mu\nu}dx^\mu dx^\nu=-\left(1+2\phi\right)dt^2+2\omega_i dx^idt+\\\left[\delta_{ij}(1-2\psi)+\chi_{ij}\right]dx^idx^j\,,\quad
\chi^{i}_{\phantom{i}i}=0\;,
\end{multline}
where the 3-vector $\omega_i$ can be decomposed into a gradient and a divergence-free part,  
\eq
\omega_i=\partial_i\omega^{\parallel}+\omega^{\bot}_i,\quad
\partial^i\omega^{\bot}_i=0\;,
\eeq
while the traceless 3-tensor $\chi_{ij}$ can be split in a gradient part, a divergence-free vector part 
and a (gauge invariant) transverse pure-tensor part,
\begin{align}
&\chi_{ij}=D_{ij}\chi^{\parallel}+\partial_{(i}\chi^{\bot}_{j)}+\chi^{\top}_{ij}\,,\; D_{ij}\equiv \partial_{i}\partial_{j}-\frac{1}{3}\delta_{ij}\nabla^2\,,\nonumber\\
&\partial^i\chi^{\bot}_{i}=\partial^i\chi^{\top}_{ij}=\chi^{\top i}_i=0\,,
\end{align}
where $\nabla^2=\delta^{ij}\partial_i\partial_j$. Note that Latin indices are raised and lowered with the Kronecker delta $\delta_{ij}$.
Similarly, the stress-energy tensor can be written as
\begin{multline}\label{eq:SE_dec}
T_{\mu\nu}dx^\mu dx^\nu=T_{tt}dt^2+2(\partial_i S^\parallel+S^{\bot}_i)dt dx^i\\+\left[\frac{T}{3}\delta_{ij}+D_{ij}\Sigma^{\parallel}+\partial_{(i}\Sigma^{\bot}_{j)}+\Sigma^{\top}_{ij}\right]dx^idx^j\,,
\end{multline}
where 
\eq\label{eq:conditions}
\partial^i S^{\bot} _i= \partial^i \Sigma^{\bot} _i=  \partial^i \Sigma^{\top}_{ij}=\Sigma^{\top i}_{i} = 0\,.
\eeq
Note that the decompositions outlined in equations \eqref{eq:metric} and \eqref{eq:SE_dec} are defined unambiguously  if the perturbations
go to zero sufficiently fast  as $r\to\infty$ so as to make the Laplacian operator $\nabla^2$ invertible. 
As an example, let us consider the case of equation \eqref{eq:SE_dec}. First, 
calculating $\partial^i T_{0i}$ and using equation \eqref{eq:conditions} one immediately obtains
\begin{gather}
 S^\parallel=\nabla^{-2}(\partial^i T_{0i})\,,\\
 S^{\bot}_i=T_{0i}-\partial_i S^\parallel\,,
\end{gather}
where $\nabla^{-2}$ denotes the inverse of the Laplacian $\nabla^2$.
Summing over the spatial indices trivially gives
\eq\label{eq:trace}
T=\delta^{ij}T_{ij}\,,
\eeq
and calculating  $\partial^i\partial^j T_{ij}$ and $\partial^j T_{ij}$ using equation \eqref{eq:conditions} one easily obtains
\begin{gather}\label{eq:par_bot}
\Sigma^{\parallel}=\nabla^{-2}\left[\frac32\nabla^{-2}\left(\partial^i\partial^j T_{ij}\right)-\frac12 T\right]\,,\\
\Sigma^{\bot}_{i}=2\nabla^{-2}\left(\partial^j T_{ij}-\frac13\partial_i T\right)-\frac43\partial_i\Sigma^{\parallel}\label{eq:par_bot2}\,.
\end{gather}
Inserting equations \eqref{eq:trace}, \eqref{eq:par_bot} and \eqref{eq:par_bot2} into equation \eqref{eq:SE_dec}, one can finally derive an explicit expression for
the gauge invariant transverse traceless perturbation $\Sigma^{\top}_{ij}$. Similar considerations apply to the decomposition  \eqref{eq:metric}
of the metric. 

We should mention that our perturbative expansion relies on \textit{two} parameters $\varepsilon_1$ and $\varepsilon_2$.
Deviations of the metric away from a flat background
are due to the presence of the fluid, which causes perturbations of dimensionless order 
$\lesssim \varepsilon_1={\cal O} ({\cal L}/\lambda_J)^2$
[$\cal L$ being the characteristic size of the medium and $\lambda_J=c_s/(4\pi(p+\rho))^{1/2}$ being a generalized Jeans length], 
and due to the presence of the perturber, which is expected to cause perturbations
of order $\varepsilon_2=M/r$, where $r$ is the distance from the perturber. 
Note that the perturbations of the first kind are small 
if the fluid is not self-gravitating (\textit{i.e.} if ${\cal L}\ll\lambda_J$),
while those of the second kind in principle diverge if we consider a point-like perturber. 
In order to retain the validity of the perturbative expansion, we therefore have to introduce a cutoff 
$r_{\min}$, which is taken to be the size of the star acting as the perturber or, in the case where the perturber is instead
a black hole, the ``capture'' impact parameter $r_{\min}\approx 2M(1+v^2)/v^2$ (\textit{i.e.,} the impact parameter for which
a test-particle is deflected by an angle $\sim 1$ by the black hole). This ensures that $\varepsilon_2$
is small and can be treated as an expansion parameter. The gravitational field produced by the perturber on scales smaller than the cutoff gives rise, when coupled to the fluid, 
to accretion onto the perturber. This gives additional contributions to the drag,
but these effects can easily be calculated separately: see for instance \citet{petrich} [equation (2.40)] for the drag-force 
due to accretion onto a black hole. When acting directly on the perturber, the gravitational field produced by the perturber itself gives rise instead to the so-called \textit{self-force} [see \citet{poisson_rev} for a review], the dissipative part of 
which accounts for the energy and angular
momentum lost through gravitational waves.

In order to exploit as much as possible the calculations done in the Newtonian case by \citet{ostriker} and \citet{circ_drag}, let us choose
the so-called Poisson gauge \citep{ma_bertschinger}, defined by the conditions $\partial_i \omega^i=\partial_i \chi^{ij}=0$. 
In this gauge the  perturbed metric is
\begin{multline}
\label{eq:metric_poisson}
d\tilde{s}^2=-\left(1+2\phi\right)dt^2+2\omega^\bot_i dx^idt+\\\left[\delta_{ij}(1-2\psi)+\chi^\top_{ij}\right]dx^idx^j\;,
\end{multline}
and  the linearized Einstein equations give
\begin{gather}
 \nabla^2 \psi =4 \pi T_{tt}\,,\label{eq:nabla_2_psi}\\
\partial_t \psi =4 \pi S^\parallel\,,\label{eq:dtpsi}\\
\nabla^2 \omega_i^{\bot} =-16 \pi S^{\bot}_i\,,\label{eq:transverse}\\
 \nabla^2 \phi =4 \pi (T_{tt}+T)-3\partial^2_t\psi\,,\label{eq:nabla_2_phi}\\
\psi-\phi=8\pi \Sigma^\parallel\,,\label{eq:psi_minus_phi}\\
\partial_t \omega^{\bot}_i=-8\pi \Sigma_i^{\bot}\,,\\
\Box \chi^\top_{ij} =-16 \pi \Sigma^{\top}_{ij}\;,\label{eq:GWs}
\end{gather}
where $\Box=\eta^{\mu\nu}\partial_\nu\partial_\mu$.
In particular, from the linearized Einstein equations one gets the following relations between the matter fields:
\begin{gather}
 \nabla^2 S^\parallel = \partial_t T_{tt}\,,\\
\nabla^2 \Sigma^\parallel =\frac12(3\partial_t S^\parallel-T)\,,\label{eq:cons2}\\
\nabla^2 \Sigma_i^\bot=2 \partial_t S^\bot_i\,,
\end{gather}
which can also be derived directly from the conservation  (to first order) of 
the stress-energy tensor with respect to the background metric, $\partial_\mu T^{\mu\nu}=0$.

Let us now write the stress energy tensor as $T_{\mu\nu}=T_{\mu\nu}^{\rm fluid}+T_{\mu\nu}^{\rm pert}$. The stress-energy tensor of the fluid is
\eq\label{eq:fluid_stress_energy}
T_{\mu\nu}^{\rm fluid}=(\tilde{p}+\tilde{\rho}) \tilde{u}_{\mu}\tilde{u}_{\nu}+\tilde{p} \tilde{g}_{\mu\nu}\,,
\eeq
where the perturbed metric $\tilde{g}_{\mu\nu}$ is given by equation \eqref{eq:metric_poisson} and $\tilde{\rho}$, $\tilde{p}$ and ${\tilde{u}}^\mu$ are 
the perturbed energy density, pressure and 4-velocity of the fluid:
\begin{gather}
 \tilde{\rho}=\rho+\delta \rho\;,\quad \tilde{p}=p+\delta p\,,\\
  \tilde {u}_i=\delta u_i\;,\quad \tilde {u}_t=-1-\phi
\end{gather}
(the equation for $\tilde {u}_i$ comes about because the fluid is at rest at the initial
position of the perturber, while the equation for $\tilde {u}_t$ follows from the 
normalization condition $\tilde{g}_{\mu\nu}\tilde{u}^\mu \tilde{u}^{\nu}=-1$).
The stress-energy tensor of the perturber is [see for instance \citet{poisson_rev}]
\begin{equation}\label{eq:pert_stress_en_full}
T^{\rm pert}_{\mu\nu}(\boldsymbol{x},t)=M \frac{\tilde{u}^{\rm pert}_\mu \tilde{u}^{\rm pert}_\nu}{\tilde{u}_{\rm pert}^t\sqrt{-\tilde{g}}}
\delta^{(3)}(\boldsymbol{x}-\tilde{\boldsymbol{x}}^{\rm pert}(t))\,,
\end{equation}
where $\tilde{u}^\mu_{\rm pert}$ and $\tilde{\boldsymbol{x}}^{\rm pert}(t)$ are the perturbed 4-velocity and spatial trajectory of the perturber, 
which for mathematical purposes is considered to be a point-particle, and $\tilde{g}$ is 
the determinant of the perturbed metric \eqref{eq:metric_poisson}.
If one expands the trajectory  $\tilde{\boldsymbol{x}}^{\rm pert}$  and the 4-velocity $\tilde{u}^\mu_{\rm pert}$ 
of the perturber as the sum of their unperturbed values $\boldsymbol{x}^{\rm pert}$ and $u^\mu_{\rm pert}$ plus some
perturbations due to the presence of the fluid (and therefore of order $\varepsilon_1$)  
and some perturbations due to the interaction of the perturber with its own gravitational field (and therefore of order $\varepsilon_2$),
and notes that $\tilde{g}= -(1+2\phi-6\psi)=-1+{\cal O} (\varepsilon_1,\varepsilon_2)$, 
equation \eqref{eq:pert_stress_en_full} can be written as
\begin{multline}\label{eq:pert_stress_en1}
T^{\rm pert}_{\mu\nu}(\boldsymbol{x},t)=M \frac{{u}^{\rm pert}_\mu {u}^{\rm pert}_\nu}{{u}_{\rm pert}^t}
\delta^{(3)}(\boldsymbol{x}-{\boldsymbol{x}}^{\rm pert}(t))\\\times [1+ {\cal O} (\varepsilon_1,\varepsilon_2)]\,.
\end{multline}
Note that because of the presence of the factor $M=r\varepsilon_2$, the stress-energy tensor $T^{\rm pert}_{\mu\nu}$ is an intrinsically first order quantity, and dropping the second order terms, as we have done earlier, we can simply write
\begin{equation}\label{eq:pert_stress_en}
T^{\rm pert}_{\mu\nu}(\boldsymbol{x},t)=M \frac{{u}^{\rm pert}_\mu {u}^{\rm pert}_\nu}{{u}_{\rm pert}^t}
\delta^{(3)}(\boldsymbol{x}-{\boldsymbol{x}}^{\rm pert}(t))\,.
\end{equation}
Perturbing the expression for the conservation of the baryon number of the fluid, 
$\partial_{\mu}((-\tilde{g})^{1/2}\,\tilde {n} \tilde {u}^\mu)=0$ ($\tilde{n}=n+\delta n$
being the perturbed number density), one gets 
\eq\label{eq:baryon}
\partial_t\left(\frac{\delta n}{n}\right)+ \partial_i \delta u^i-3\partial_t\psi=0\,,
\eeq
whereas perturbing the Euler equation $\tilde {a}^\mu=\tilde{u}^\alpha\tilde{\nabla}_\alpha \tilde{u}^\mu=-(\tilde{g}^{\mu\nu}+\tilde{u}^\mu \tilde{u}^{\nu})\partial_\nu \tilde{h}/\tilde{h}$ [$\tilde{h}\equiv(\tilde{p}+\tilde{\rho})/\tilde{n}=h+\delta h$ is the perturbed specific enthalpy] one obtains
\eq\label{eq:euler}
\partial_t \delta u^i+ \partial_i\phi +\partial_t\omega^\bot_i+c_s^2\partial_i\frac{\delta n}{n}=0\,,
\eeq
where $c_s=(dp/d\rho)^{1/2}$ is the velocity of sound and where we have used the first law of thermodynamics ($\delta h/h=c_s^2 \delta n/n$).
Combining the divergence of equation \eqref{eq:euler} and the time derivative of equation \eqref{eq:baryon} and finally using equation \eqref{eq:nabla_2_phi}, one gets the following wave equation for the baryon density perturbations:
\eq\label{eq:delta_n_wave}
(\partial_t^2-c_s^2\nabla^2)\frac{\delta n}{n}=\nabla^2 \phi+3\partial^2_t\psi=4 \pi (T_{tt}+T)\;.
\eeq

 In the next sections we will solve  the wave equation \eqref{eq:delta_n_wave} and the Einstein equations \eqref{eq:nabla_2_psi}-\eqref{eq:GWs}. It is understood, however, that since these equations are linearized, the solutions that we find ``intrinsically'' have a relative error ${\cal O}(\varepsilon_1,\varepsilon_2)$. This error is not to be confused with those which we will introduce when solving these equations approximately. We will explicitly keep track of the latter in the next sections, while we will re-introduce the relative error ${\cal O}(\varepsilon_1,\varepsilon_2)$ due to the linearization procedure only in the final results.

\section{\label{sec:straight}Straight-line motion}
%
Let us first consider the case of a perturber moving along a straight-line, which is taken to be the $z$-axis of a Cartesian 
coordinate system: the  unperturbed trajectory of the perturber is therefore 
$x^{\rm pert}(t)=y^{\rm pert}(t)=0$, $z^{\rm pert}(t)=v t$ and the  unperturbed 4-velocity reads
$u_{\rm pert}^\mu \partial/\partial x^\mu=\gamma(\partial/\partial t+v \partial/\partial z)$, with $\gamma^2=1/(1-v^2)$.
Denoting by $H(t)$ the step function, equation \eqref{eq:delta_n_wave} can be rewritten as
\begin{multline}\label{eq:wave}
(\partial_t^2-c_s^2\nabla^2)\frac{\delta n}{n}=4\pi M\gamma(1+v^2)\delta(x)\delta(y)\delta(z-v t)H(t)\\+4 \pi [\rho(1+2\phi)+3p(1-2\psi)]+4 \pi (1+3 c_s^2)(p+\rho)\frac{\delta n}{n}\,.
\end{multline}
Solving this equation is complicated by the presence of the terms $4 \pi [\rho(1+2\phi)+3p(1-2\psi)]$ and 
$4 \pi (1+3 c_s^2)(p+\rho){\delta n}/{n}$ on the right-hand side. If these terms were not present, we could simply solve equation
\eqref{eq:wave} by using the Green's function of the flat wave operator $-\partial_t^2+c_s^2\nabla^2$, and proceeding as in
\citet{ostriker} we would get 
\begin{align}\label{eq:baryon_sol_simplified}
\frac{\delta n}{n}(\boldsymbol{x},t)\approx f\frac{M\gamma(1+v^2)}{c_s^2[(z-v t)^2+(x^2+y^2)(1-{\cal M}^2)]^{1/2}}\;,
\end{align}
where ${\cal M}=v/c_s$ is the Mach number and
\begin{equation}
f=\left\{
\begin{array}{l}
1\quad \mbox{if $x^2+y^2+z^2< (c_st)^2$}\\\\
2\quad \mbox{if ${\cal M}>1$, $x^2+y^2+z^2> (c_st)^2$,}\\
\mbox{$(z-v t)/\sqrt{x^2+y^2}<-\sqrt{{\cal M}^2-1}$}\\\mbox{and $z>c_st/{\cal M}$}\\\\
0\quad \mbox{otherwise}
\end{array}\right.
\end{equation}
Note that performing a boost to 
the reference frame comoving with the perturber (the ``primed'' frame) this ``approximate'' solution becomes 
\begin{equation}\label{eq:comparison}
\frac{\delta n}{n}\approx f\frac{M\gamma^2(1+v^2)}{c_s^2r'\sqrt{1-{\cal{\tilde{M}}}^2\sin^2\theta'}}\,,
\quad{\cal{\tilde{M}}}^2=\frac{1-c_s^2}{c_s^2}\gamma^2v^2\,,
\end{equation}
where $r'$ and $\theta'$ are the radius and polar angle in the primed frame \textit{i.e.,}
$r'=\sqrt{x'^2+y'^2+z'^2}$ and $\cos\theta'=z'/r'$ in terms of the Cartesian coordinates $x'$, $y'$ and $z'$.
Equation \eqref{eq:comparison} agrees with the solution found 
in \citet{petrich} [equation (B30)], except for the different value of $f$ [this happens because \citet{petrich} considered the stationary solution instead of performing a finite-time analysis: \textit{cf.} \citet{ostriker} for more details].

It is not difficult to see that equations \eqref{eq:baryon_sol_simplified} and \eqref{eq:comparison} are actually approximate solutions to
equation \eqref{eq:wave}. Indeed, the term $4 \pi (\rho+3p)$ on the right-hand side of equation \eqref{eq:wave} simply gives rise
to an error ${\cal O}({\cal L}/\lambda_J)^2$ in the solution. This error represents the 
correction due to the fact that having a fluid with constant $p$ and $\rho$
together with the Minkowski metric is not a solution of the Einstein equations. When it comes to the term
$8 \pi (\rho\phi+3p\psi)$, let us note that the gravitational potentials $\phi$ and $\psi$ 
consist of a part of order ${\cal O}({\cal L}/\lambda_J)^2$ due to the presence of the fluid and a part of order $-M/r\sim-\delta n/n$ ($r$ being the
distance from the perturber) due to the presence of the perturber.
The first part of the potentials therefore gives rise, when inserted into the  term
$8 \pi (\rho\phi+3p\psi)$, to an error much smaller than the error ${\cal O}({\cal L}/\lambda_J)^2$ coming from the term $4 \pi (\rho+3p)$.
The second part of the potentials, when inserted into the term
$8 \pi (\rho\phi+3p\psi)$, gives rise instead to a Yukawa-like term similar to the term $4 \pi (1+3 c_s^2)(p+\rho){\delta n}/{n}$
appearing on the right-hand side of equation \eqref{eq:wave}. It is not difficult to see that these Yukawa-like terms give rise to a relative error $\varepsilon_{\rm J}\sim{\cal O}({\cal L}/\lambda_J)$. To see this, one can simply Fourier-transform equation  \eqref{eq:wave} with respect to time
in order to get rid of the time derivatives. One is then left with an equation of the form
\eq\label{eq:cartoon2}
\nabla^2\left(\frac{\delta n}{n}(\boldsymbol{x},\omega)\right)\approx S(\boldsymbol{x},\omega)
+(a/\lambda_{\rm J}^2-\omega^2/c_s^2)\frac{\delta n}{n}(\boldsymbol{x},\omega)\;,
\eeq
where for simplicity we have used the same symbol for $\delta n/n$ and its Fourier transform with respect to time, $a$ is a constant and  
$S(\boldsymbol{x},\omega)$ is a suitably defined source function [inspection of equation \eqref{eq:wave} actually reveals that  $S(\boldsymbol{x},\omega)\sim \exp(i\omega z/v)$].
Using the Green's function of the Yukawa operator $\nabla^2-\mu^2$ ($\mu$ being a constant),\footnote{Using spherical
coordinates and the fact that $\nabla^2(1/\vert\boldsymbol{x}\vert)=-4\pi\delta^{(3)}(\boldsymbol{x})$, it is indeed easy to
check that $(\nabla^2-\mu^2)G(\boldsymbol{x})=-\delta^{(3)}(\boldsymbol{x})$.}
\eq\label{eq:green}
G(\boldsymbol{x})=\frac{\exp{(-\mu \vert\boldsymbol{x}\vert)}}{4\pi\vert\boldsymbol{x}\vert}\,,
\eeq
this equation can be solved and gives
\begin{multline}\label{eq:cartoon_sol}
\frac{\delta n}{n}(\boldsymbol{x},\omega)=\\-\int d^3x'
\frac{\exp{(-\sqrt{a/\lambda_{\rm J}^2-\omega^2/c_s^2}\vert\boldsymbol{x}-\boldsymbol{x}'\vert)}}{4\pi\vert\boldsymbol{x}-\boldsymbol{x}'\vert}
S(\boldsymbol{x}',\omega)\;.
\end{multline}
If $\omega/c_s\gg 1/\lambda_{\rm J}$, 
one can series expand equation \eqref{eq:cartoon_sol} and get
 \begin{align}\label{eq:cartoon_sol2}
 &\frac{\delta n}{n}(\boldsymbol{x},\omega)\approx\\
 &-\int d^3x'\Big\{\frac{\exp{[i a c_s\vert\boldsymbol{x}-\boldsymbol{x}'\vert /(2\omega\lambda_{\rm J}^2)]}}{4\pi\vert\boldsymbol{x}-\boldsymbol{x}'\vert}\nonumber\\
 &\qquad\qquad\qquad\qquad\times\exp(-i\omega/c_s\vert\boldsymbol{x}-\boldsymbol{x}'\vert)S(\boldsymbol{x}',\omega)\Big\}\approx
 \nonumber\\ &-\int d^3x'\frac{\exp(-i\omega/c_s\vert\boldsymbol{x}-\boldsymbol{x}'\vert)S(\boldsymbol{x}',\omega)}{4\pi\vert\boldsymbol{x}-\boldsymbol{x}'\vert}
 \times\left(1+\frac{c_s\varepsilon_{\rm J}}{\omega\lambda_{\rm J}}\right)\nonumber
 \;,
 \end{align}
and from the last line of this equation it is clear that one gets the solution which would have been obtained by neglecting the term $a/\lambda_{\rm J}^2 \times{\delta n}/{n}$ in equation \eqref{eq:cartoon2}, with a relative error
$c_s\varepsilon_{\rm J}/(\omega\lambda_{\rm J})\ll\varepsilon_{\rm J}$.
For frequencies  $\omega/c_s\ll 1/\lambda_{\rm J}$ (\textit{i.e.,} for wavelengths larger than the generalized Jeans length $\lambda_{\rm J}$) this procedure is not applicable. 
However, it is clear that for $\omega=0$ equation \eqref{eq:cartoon_sol}
becomes the solution which would have been obtained by neglecting the term $a/\lambda_{\rm J}^2 \times {\delta n}/{n}$ in equation \eqref{eq:cartoon2}, corrected by a factor $\sim(1+\varepsilon_{\rm J})$. Moreover, because $S(\boldsymbol{x},\omega)\sim \exp(i\omega z/v)$, the integral appearing in equation  \eqref{eq:cartoon_sol} averages out if $\omega\gg v/{\cal L}$. Therefore, the spectrum of $\delta n/n$
extends up to $\omega_{\rm cutoff}\sim v/{\cal L}$, and the effect of the frequencies 
$\omega/c_s\ll 1/\lambda_{\rm J}$ on the final solution $\delta n/n(\boldsymbol{x},t)$ is negligible because 
$\omega_{\rm cutoff}\sim v/{\cal L}\gg 1/\lambda_{\rm J}$
if the fluid is not self-gravitating.

As such, since we already know the solution of equation \eqref{eq:wave} if we neglect the terms $4 \pi [\rho(1+2\phi)+3p(1-2\psi)]$ and 
$4 \pi (1+3 c_s^2)(p+\rho){\delta n}/{n}$ on the right-hand side [equation \eqref{eq:baryon_sol_simplified}], we can write the following approximate solution for equation \eqref{eq:wave}:
\begin{align}\label{eq:baryon_sol}
\frac{\delta n}{n}(\boldsymbol{x},t)&\,=f\frac{M\gamma(1+v^2)}{c_s^2[(z-v t)^2+(x^2+y^2)(1-{\cal M}^2)]^{1/2}}\nonumber\\
&\times(1+\varepsilon_{\rm J})+{\cal O}({\cal L}/\lambda_J)^2\;,
\end{align}
where, as explained above, the error ${\cal O}({\cal L}/\lambda_J)^2$ comes from the term $4 \pi (\rho+3p)$ on the right-hand side of equation \eqref{eq:wave}, while the error $\varepsilon_{\rm J}$ comes from the terms $4 \pi (1+3 c_s^2)(p+\rho){\delta n}/{n}$ and $8 \pi (\rho\phi+3p\psi)$.
[Note that $\varepsilon_{\rm J}(x,y,z,t)=\varepsilon_{\rm J}(-x,y,z,t)$ and  $\varepsilon_{\rm J}(x,y,z,t)=\varepsilon_{\rm J}(x,-y,z,t)$
due to the cylindrical symmetry of the problem.] 
Both of these errors are negligible if ${\cal L}\ll \lambda_J$ (\textit{i.e.,} if the fluid is not self-gravitating). 

The trajectory of the perturber is governed by the geodesic equation of the physical, perturbed spacetime (\textit{i.e.,} the one with metric $\tilde{g}_{\mu\nu}=\eta_{\mu\nu}+\delta g_{\mu\nu}$). The familiar form of this equation is
\eq
\frac{d^2\tilde{x}_{\rm pert}^\mu}{d\tilde{\tau}^2}+\tilde{\Gamma}^\mu_{\alpha\beta}\frac{d\tilde{x}_{\rm pert}^\alpha}{d\tilde{\tau}}\frac{d\tilde{x}_{\rm pert}^\beta}{d\tilde{\tau}}=0\,,
\eeq
where $\tilde{x}_{\rm pert}^\mu$ and $\tilde{\tau}$ are the perturbed trajectory and proper time while 
the $\tilde{\Gamma}$'s are the Christoffel symbols of the perturbed spacetime. This equation can be easily expressed in
terms of the background proper time $\tau$, 
\eq
\frac{d^2\tilde{x}_{\rm pert}^\mu}{d{\tau}^2}+\tilde{\Gamma}^\mu_{\alpha\beta}\frac{d\tilde{x}_{\rm pert}^\alpha}{d{\tau}}\frac{d\tilde{x}_{\rm pert}^\beta}{d{\tau}}
=-\frac{d^2\tau}{d\tilde{\tau}^2}\left(\frac{d\tilde{\tau}}{d\tau}\right)^2\frac{d\tilde{x}_{\rm pert}^\mu}{d\tau}\,,
\eeq
which can be also written as
\begin{multline}\label{eq:gen_geo}
\frac{d^2\tilde{x}_{\rm pert}^\mu}{d{\tau}^2}+\tilde{\Gamma}^\mu_{\alpha\beta}\frac{d\tilde{x}_{\rm pert}^\alpha}{d{\tau}}\frac{d\tilde{x}_{\rm pert}^\beta}{d{\tau}}=\\
\frac{{d\tilde{x}_{\rm pert}^\mu}/{d\tau}}{\sqrt{-\tilde{g}_{\alpha\beta}\frac{d\tilde{x}_{\rm pert}^\alpha}{d\tau}\frac{d\tilde{x}_{\rm pert}^\beta}{d\tau}}}\frac{d}{d\tau}
\sqrt{-\tilde{g}_{\alpha\beta}\frac{d\tilde{x}_{\rm pert}^\alpha}{d\tau}\frac{d\tilde{x}_{\rm pert}^\beta}{d\tau}}
\,.
\end{multline}
Using now $\tilde{g}_{\mu\nu}=\eta_{\mu\nu}+\delta g_{\mu\nu}$, equation \eqref{eq:gen_geo}
can be easily rewritten, to first order [\textit{i.e.,} neglecting as usual errors 
of order ${\cal O}(\varepsilon_1^2,\varepsilon_2^2,\varepsilon_1\varepsilon_2)$], as \citep{poisson_rev}
\begin{multline}\label{eq:acceleration}
\tilde{a}_{\rm pert}^\mu=\frac{d^2\tilde{x}_{\rm pert}^\mu}{d{\tau}^2}=\\-\frac12(\eta^{\mu\nu}+u_{\rm pert}^\mu u_{\rm pert}^\nu)(2 \partial_\rho \delta g_{\nu\lambda}-\partial_\nu \delta g_{\rho\lambda})u_{\rm pert}^\lambda u_{\rm pert}^\rho\,.
\end{multline}
The metric perturbations $\delta g_{\mu\nu}$ appearing on the right-hand side of equation \eqref{eq:acceleration} consist of a part produced by the stress-energy of the fluid ($\delta g_{\mu\nu}^{\rm F}$) and  
one produced by the perturber ($\delta g_{\mu\nu}^{\rm P}$). The latter contribution,
as already mentioned, gives rise to accretion onto the perturber and to the self-force.
The drag due to accretion is easy to calculate separately, as mentioned previously, while the self-force is
in general hard to deal with \citep{poisson_rev}.
However, it is well-known that the self-force is zero in a Minkowski spacetime for \textit{geodetic} (\textit{i.e.,} 
straight-line) motion \textit{in the Lorenz gauge}. Since the right-hand side of equation \eqref{eq:acceleration} is not gauge-invariant, the self-force itself is \textit{not} gauge-invariant \citep{gauge}. Nevertheless, it is possible to show that at least the dissipative part of the self-force (\textit{i.e.,} the one accounting for the deceleration due to the loss of energy and angular momentum through gravitational waves) is gauge-invariant and therefore zero also in the gauge which we are using \citep{mino}. (Alternatively, this can be understood from the fact that a perturber moving on a straight-line does 
not emit energy through gravitational waves in the quadrupole approximation.) 
It should be noted that the presence of the fluid does not 
alter these results. In fact, one can insert the decomposition $\delta g_{\mu\nu}=\delta g_{\mu\nu}^{\rm F}+\delta g_{\mu\nu}^{\rm P}$
into the Einstein equations, and split them into equations for $\delta g_{\mu\nu}^{\rm F}$ 
and equations for $\delta g_{\mu\nu}^{\rm P}$ by including in the right-hand sides of the equations for $\delta g_{\mu\nu}^{\rm P}$ 
only quantities containing the stress-energy of the perturber and $\delta g_{\mu\nu}^{\rm P}$ itself. In particular from equations \eqref{eq:nabla_2_psi}, \eqref{eq:nabla_2_phi}
and \eqref{eq:GWs}, using equation \eqref{eq:fluid_stress_energy} one gets 
\begin{gather}
 \nabla^2 \psi^{\rm P} =8\pi\rho \phi^{\rm P}+4 \pi T^{\rm pert}_{tt}\,,\label{eq:nabla_2_psi_P}\\
 \nabla^2 \psi^{\rm F} =  4 \pi \left[\rho+ 2 \rho \phi^{\rm F}+ (p+\rho) \frac{\delta n}{n}\right]\,,\label{eq:poisson_eq}\\
 \psi^{\rm P}-\phi^{\rm P}=8\pi \Sigma^\parallel_{\rm pert}\label{eq:psi_minus_phi_P}\,,\\
 \psi^{\rm F}-\phi^{\rm F}=0\label{eq:psi_minus_phi_F}\,,\\
 \Box \chi_{ij}^{\top\,{\rm P}} =-16 \pi (\Sigma^{\top \,{\rm pert}}_{ij}+p \chi_{ij}^{\top\,{\rm P}} )\,,\label{eq:GWs_P}\\
 \Box \chi_{ij}^{\top\,{\rm F}} =-16 \pi p \chi_{ij}^{\top\,{\rm F}} \;.\label{eq:GWs_F}
\end{gather}
From equation \eqref{eq:euler}  
it follows instead that $\delta u_i^\bot=-\omega_i^\bot$ \footnote{We are making here the simplifying but reasonable assumption
that no vortical modes $\delta u_i^\bot$ and $\omega_i^\bot$ are excited in the system before the perturber is turned on at $t=0$.} 
and therefore $S_i^{\bot\,{\rm fluid}}=-(p+\rho)\delta u_i^\bot+p\omega_i^\bot=(2p+\rho)\omega_i^\bot$, 
which together with equation \eqref{eq:transverse} gives
\begin{gather}
\nabla^2 \omega_{i}^{\bot\,{\rm P}} =-16 \pi S^{\bot\,{\rm pert}}_{i}-16 \pi (2 p+\rho)\omega_{i}^{\bot\,{\rm P}}\,,\label{eq:transverse_P}\\
\nabla^2 \omega_{i}^{\bot\,{\rm F}} =-16 \pi (2 p+\rho)\omega_{i}^{\bot\,{\rm F}}\label{eq:transverse_F}\;.
\end{gather}
From equations \eqref{eq:nabla_2_psi_P}, \eqref{eq:psi_minus_phi_P}, \eqref{eq:GWs_P} and 
\eqref{eq:transverse_P} it therefore follows that the metric perturbations $\delta g_{\mu\nu}^{\rm P}$ 
produced by the perturber are the same as in the absence of the fluid,
except for the presence of the terms $8\pi\rho \phi^{\rm P}$, $-16 \pi p \chi^{\top\,{\rm P}}_{ij}$ and
$-16 \pi (2 p+\rho)\omega_i^{\bot\,{\rm P}}$ on the right-hand sides of equations \eqref{eq:nabla_2_psi_P},  
\eqref{eq:GWs_P} and \eqref{eq:transverse_P}. 
Using the Green's function of the Yukawa operator 
it is easy to see that
these terms produce a contribution of order ${\cal O}(\rho\, r_{\min}^2\,\partial\delta 
g_{\mu\nu}^{\rm P})\sim{\cal O}(\rho\, M)$ to the gradients $\partial\delta g_{\mu\nu}^{\rm P}\sim M/r_{\min}^2$. 
To be more specific, let us consider for example the case of $\psi^{\rm P}$.
Using equations \eqref{eq:green}, \eqref{eq:nabla_2_psi_P} and \eqref{eq:psi_minus_phi_P}, 
the solution for $\psi^{\rm P}$ reads
\begin{align}
&\psi^{\rm P}(\boldsymbol{x},t)=\\&\!\!-\!\!\int\! d^3x'
\frac{\exp{(-\sqrt{8\pi\rho}\vert\boldsymbol{x}-\boldsymbol{x}'\vert)}}{\vert\boldsymbol{x}-\boldsymbol{x}'\vert}
\!\left[ T^{\rm pert}_{tt}(\boldsymbol{x}',t)\!-\!16\pi\rho \Sigma^\parallel_{\rm pert}(\boldsymbol{x}',t)\right].\nonumber
\end{align}
Taking now the derivative with respect to $\boldsymbol{x}$ and expanding the exponential, it is easy to check that the 
 presence of the fluid simply adds a contribution of order ${\cal O}(\rho\, r_{\min}^2\,\partial_i\psi^{\rm P})$ 
to the gradient $\partial_i\psi^{\rm P}$. It should be noted that a contribution of order ${\cal O}(\rho\, r_{\min}^2\,\partial\delta
g_{\mu\nu}^{\rm P})\sim{\cal O}(\rho\, M)$ to the gradients $\partial\delta g_{\mu\nu}^{\rm P}$ corresponds to
a contribution of order ${\cal O}(\rho\, M^2)$ to the drag: this contribution can be interpreted, 
as we have mentioned, as being due to accretion onto the perturber. 

We will therefore focus on the force produced by the gravitational
interaction with the fluid, which includes dynamical friction.
From equations \eqref{eq:poisson_eq}, \eqref{eq:psi_minus_phi_F}, \eqref{eq:GWs_F} and \eqref{eq:transverse_F}
it follows that the fluid can only excite the metric perturbations $\phi$ and $\psi$.
Using equations \eqref{eq:green}, \eqref{eq:poisson_eq} and \eqref{eq:psi_minus_phi_F},  
we can easily get expressions for the gradients $\partial_\mu\phi^{\rm F}=\partial_\mu\psi^{\rm F}$ 
evaluated at the position of the perturber $x=y=0$, $z=v t$, which enter equation \eqref{eq:acceleration}.
In particular, the solution for $\psi^{\rm F}$ is
\begin{multline}\label{eq:psi_expl}
\psi^{\rm F}(\boldsymbol{x},t)=\\-\int d^3x'
\frac{\exp{(-\sqrt{8\pi\rho}\vert\boldsymbol{x}-\boldsymbol{x}'\vert)}}{\vert\boldsymbol{x}-\boldsymbol{x}'\vert}
\left[ \rho+ (p+\rho)\frac{\delta n}{n}(\boldsymbol{x}',t)\right]\,,
\end{multline}
and taking the derivative with respect to $\boldsymbol{x}$, one easily gets
\begin{multline}\label{eq:psi_grad_explicit}
\partial_i \psi^{\rm F}(\boldsymbol{x})=\\\int d^3x'
\frac{x^i-{x^i}'}{\vert\boldsymbol{x}-\boldsymbol{x}'\vert^3}
\left[ \rho+ (p+\rho)\frac{\delta n}{n}(\boldsymbol{x}',t)\right]\times(1+\varepsilon_{\rm Yukawa})\,,
\end{multline}
where we have introduced the error $\varepsilon_{\rm Yukawa}\sim{\cal O} (\rho{\cal L}^2)$ which arises when expanding the Yukawa exponential.
In particular, note that the source $\rho$ appearing in the integral of equation \eqref{eq:psi_grad_explicit} simply gives
the gravitational force exerted by the unperturbed medium
on the perturber. This force is exactly zero if the medium is spherically symmetric with respect to the perturber,
but in general the net effect on the gradients $\partial_i\phi^{\rm F}=\partial_i\psi^{\rm F}$
can be non-zero and at most of order $\rho\,{\cal L}$, depending on the shape of the fluid configuration and on
the position of the perturber. 
Similarly, the term $(p+\rho) {\delta n}/n$ appearing in equation \eqref{eq:psi_grad_explicit} can be considered as the sum of two parts,
one coming from the error
${\cal O}({\cal L}/\lambda_J)^2$ appearing in equation \eqref{eq:baryon_sol} and the other one from the rest of this equation.
Note that the first part is present even if the mass of the perturber goes to zero and
represents the force exerted by the density perturbations which appear because, as mentioned earlier, a fluid with constant $p$ and $\rho$
together with the Minkowski metric is not a solution of the Einstein equations. The contribution to the gradients 
$\partial_\mu\phi^{\rm F}=\partial_\mu\psi^{\rm F}$ from this term can be as large
as $\rho{\cal L}({\cal L}/\lambda_J)^2$, and in what follows we will group it together with the contribution from the term $\rho$ appearing in
equation \eqref{eq:psi_grad_explicit} into a correction $\varepsilon_{\rm not\,DF}\lesssim {\cal O}(\rho\,{\cal L})$.
The rest of the term $(p+\rho) {\delta n}/n$ gives instead the force exerted 
by the density perturbations produced by the perturber \textit{i.e.,} dynamical friction.
In particular, using equation \eqref{eq:baryon_sol} in  equation \eqref{eq:psi_grad_explicit}
one obtains, for the $x$ and $y$ gradients evaluated at the position of the perturber $x=y=0$, $z=v t$,
\begin{equation}
\label{eq:grad_xy} \partial_x\phi^{\rm F}=\partial_x\psi^{\rm F}=\partial_y\phi^{\rm F}=\partial_y\psi^{\rm F}=\varepsilon_{\rm not\,DF}\,,
\end{equation}
as expected from the cylindrical symmetry of the problem,
while the $t$ and $z$ gradients, evaluated at the position of the perturber $x=y=0$, $z=v t$, are
\begin{align}
\label{eq:grad_zt}&\partial_z\psi^{\rm F} =\partial_z\phi^{\rm F} = - \frac{\partial_t\psi^{\rm F}}{v}= - \frac{\partial_t\phi^{\rm F}}{v} =\nonumber
\\&(p+\rho)\int d^3{x}' \frac{\frac{\delta n}{n}\left(\boldsymbol{x}',t=\frac{z}{v}\right) (z-z')}{[x'^2+y'^2+(z'-z)^2]^{3/2}}\nonumber
\\&\times(1+\varepsilon_{\rm Yukawa})
+\varepsilon_{\rm not \, DF}\,.
\end{align}
[Note that this expression for $\partial_t\psi^{\rm F}$ is obtained by taking the derivative of equation \eqref{eq:psi_expl}
with respect to $t$, transforming the derivative with respect to $t$ acting on $\delta n/n$ into a derivative with respect to $z'$ 
using equation \eqref{eq:baryon_sol}, integrating by parts and finally transforming the derivative with respect to $z'$ into one with respect to $z$.]
The integral in equation \eqref{eq:grad_zt} can be evaluated using equation \eqref{eq:baryon_sol} as in \citet{ostriker}, and is
\begin{align}\label{eq:force}
&\partial_z\psi^{\rm F} = \partial_z\phi^{\rm F} = - \frac{\partial_t\psi^{\rm F}}{v}= - \frac{\partial_t\phi^{\rm F}}{v} =\\
&\frac{4 \pi (p+\rho) M \gamma (1+v^2)}{v^2}\, I\times[1+{\cal O} ({\cal L}/\lambda_J)]+\varepsilon_{\rm not\,DF}\,,\nonumber 
\end{align}
where 
\begin{equation}\label{eq:I}
I=\left\{
\begin{array}{ll}
 \frac12 \ln\left(\frac{1+{\cal M}}{1-{\cal M}}\right)-{\cal M}&\mbox{if ${\cal M}<1$}\,,\\
 \frac12 \ln\left(1-\frac{1}{{\cal M}^2}\right)+\ln\left(\frac{v t}{r_{\rm min}}\right)&\mbox{if ${\cal M}>1$}
\end{array}
\right.
\end{equation}
and we have made the assumptions that $|c_s-v|t$ exceeds the cutoff $r_{\min}$
and that $|c_s+v|t$ is smaller than ${\cal L}$. 

Inserting equation \eqref{eq:grad_xy} into equation \eqref{eq:acceleration}, one immediately finds  
\eq \left(\tilde{a}^x_{\rm pert}\right)_{\rm F}=\left(\tilde{a}^y_{\rm pert}\right)_{\rm F}=\varepsilon_{\rm not\,DF}\,,\eeq
while using equation \eqref{eq:force} in equation \eqref{eq:acceleration} gives
\begin{align}\label{eq:at}
 \left(\tilde{a}^t_{\rm pert}\right)_{\rm F}=\,&-\frac{4 \pi (p+\rho) M \gamma^3 (1+v^2)^2}{v} I
  \\&\times[1+{\cal O} ({\cal L}/\lambda_J)+{\cal O} (M/r_{\min})]+\varepsilon_{\rm not\,DF}\,,\nonumber\\\label{eq:az}
  \left(\tilde{a}^z_{\rm pert}\right)_{\rm F}=\,&-\frac{4 \pi (p+\rho) M \gamma^3 (1+v^2)^2}{v^2} I\\&\times[1+{\cal O} ({\cal L}/\lambda_J)+{\cal O} (M/r_{\min})]+\varepsilon_{\rm not\,DF}\,,\nonumber
\end{align}
where $I$ is defined by equation \eqref{eq:I}. Note that we have restored the relative error ${\cal O} (\varepsilon_1,\varepsilon_2)$ due to the linearization of the equations of the previous section: this gives rise to the error ${\cal O} (M/r_{\min})$ appearing in equations \eqref{eq:at} and \eqref{eq:az}.

Performing a boost we can calculate the change of 3-momentum in the rest frame of the perturber 
due to the gravitational interaction with the fluid, so as to compare with the results of \citet{petrich}:
\begin{align}\label{eq:z_drag}
 &\left(\frac{d \tilde{p}^{(z)}_{\rm pert}}{d\tau}\right)_{\rm F}=M\gamma\left[\left(\tilde{a}^z_{\rm pert}\right)_{\rm F}-v \left(\tilde{a}^t_{\rm pert}\right)_{\rm F}\right]=\\&-\frac{4 \pi (p+\rho) M^2\gamma^2 (1+v^2)^2}{v^2}\, I\times
 [1+{\cal O} ({\cal L}/\lambda_J)+{\cal O} (M/r_{\min})]\nonumber\\&+\varepsilon_{\rm not\, DF}\,,\nonumber\\
&\left(\frac{d \tilde{p}^{(x)}_{\rm pert}}{d\tau}\right)_{\rm F}=\left(\frac{d \tilde{p}^{(y)}_{\rm pert}}{d\tau}\right)_{\rm F}=\varepsilon_{\rm not\, DF}\,.
\end{align}
Note that the relative errors ${\cal O}({\cal L}/\lambda_J)$ 
and ${\cal O} (M/r_{\min})$ are negligible 
-- the former because the fluid is not self-gravitating and the latter
because the effective cutoff radius $r_{\min}$ is large compared with $M$ -- whereas 
$\varepsilon_{\rm not\, DF}$ in general is not negligible.
However, $\varepsilon_{\rm not\, DF}$ represents the standard
force acting on the perturber because of the gravitational interaction with the fluid, and it can be computed separately if the global structure of the
system is known. In particular, $\varepsilon_{\rm not\, DF}=0$ if the medium is distributed in a spherically symmetric fashion around the perturber.

\begin{figure}
\includegraphics[width=6.1cm, angle=-90]{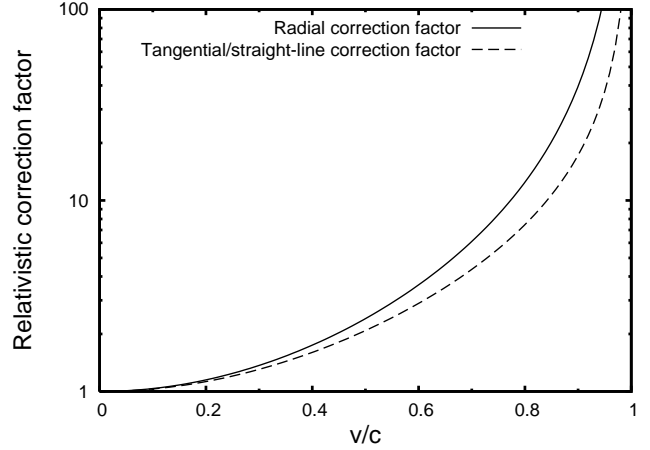}
\caption{The relativistic correction factors $\gamma^2[1+(v/c)^2]^2$, multiplying the Newtonian drag for straight-line motion 
and the tangential Newtonian drag for circular motion, and $\gamma^3[1+(v/c)^2]^2$, multiplying the radial Newtonian drag for circular motion,
are plotted as functions of the velocity $v$ of the perturber relative to the fluid. Note that velocities $v\sim0.8$ can be obtained for a 
perturber orbiting around an accreting SMBH in the opposite direction with respect to the accretion flow \citep{nextpaper}.
\label{fig:corrections}}
\end{figure}

The relativistic correction factor $\gamma^2 (1+v^2)^2$ appearing in equation \eqref{eq:z_drag}
is plotted as a function of the velocity $v$ in Fig. \ref{fig:corrections}.
Note that, for ${\cal M}\gg1$ and $v t\to r_{\rm max}$, equation \eqref{eq:z_drag}, 
and in particular the correction factor,  agrees with equation (B45) of \citet{petrich}. 
%
\section{\label{sec:circular}Circular motion}
%
Let us now consider the case of a perturber moving on a circular orbit of radius $R$ with velocity $v=\Omega R$.
Such an orbit is clearly not allowed in a Minkowski background, unless there is an external \textit{non-gravitational} force keeping the 
perturber on a circular trajectory. In astrophysical scenarios 
we are interested instead in a perturber maintained in 
circular motion by \textit{gravitational} forces. In this case, the background spacetime is necessarily curved: 
one can think of a circular orbit around a Schwarzschild black hole with mass $M_{\rm BH}$ surrounded by a tenuous fluid at rest.
However, if the perturber is sufficiently far from the central black hole (\textit{i.e.,} if $R\gg M_{\rm BH}$) one can approximately
consider the metric as given by equation \eqref{eq:metric_poisson} 
(\textit{i.e.,} Minkowski plus the perturbations produced by the presence of the fluid and of the perturber) 
and neglect the corrections ${\cal O}(M_{\rm BH}/R)$ due to the presence of the central black hole. 
This treatment is clearly not completely satisfactory, because orbital velocities become relativistic only close to the central black hole 
[in fact, $v\sim (M_{\rm BH}/R)^{1/2}$], but we argue that it may not be such a bad approximation as it might seem.

Indeed, if one uses Fermi normal coordinates comoving with the perturber [see for instance \citet{gravitation}], 
all along the trajectory the metric can be written as Minkowski plus perturbations produced
by the fluid and the perturber, the curvature of the background introducing just 
corrections of order ${\cal O}(r/M_{\rm BH})^2$ ($r$ being the \textit{spatial} distance from
the perturber). Because the wake can extend out to distances of order $R$ from the perturber \citep{circ_drag}, it will eventually feel
the curvature of the background unless $R\ll M_{\rm BH}$. However, the part of the wake giving the largest gravitational attraction  
to the perturber will be the closest to it, and this part will experience an approximately flat spacetime.  
Reasoning in the same way, we can argue that our treatment should be approximately applicable also to a perturber moving
on a circular orbit in a fluid which is moving circularly in the same plane as the perturber (\textit{e.g.,} a perturber moving inside
an accretion disc), provided that the velocity $v=\Omega R$ of the perturber is taken to be the velocity \textit{relative} to the fluid.

Considering therefore a Minkowski background spacetime, one can proceed as in the previous section, and equation \eqref{eq:delta_n_wave} becomes 
\begin{align}\label{eq:wave_circ}
&(\partial_t^2-c_s^2\nabla^2)\frac{\delta n}{n}=\frac{4\pi M\gamma(1+v^2)}{R}\delta(r-R)\delta(z)\delta(\theta-\Omega t)H(t)\nonumber\\&+4 \pi [\rho(1+2\phi)+3p(1-2\psi)]+4 \pi (1+3 c_s^2)(p+\rho)\frac{\delta n}{n}\,,
\end{align}
where we have introduced a system of cylindrical coordinates $(r,\theta,z)$ such that the motion of the perturber takes place at $z=0$, $r=R$.
The solution to this equation is rather complex, but has fortunately been worked out by \citet{circ_drag}. For our purposes, proceeding as in the previous section we can simply write it as
\begin{equation}\label{eq:baryon_sol_circ}
\frac{\delta n}{n}(\boldsymbol{x},t)=\frac{M\gamma(1+v^2)}{R\,c_s^2}{\cal D}(\boldsymbol{x},t)\times(1+\varepsilon_{\rm J})
+{\cal O}({\cal L}/\lambda_J)^2\;,
\end{equation}
where $\varepsilon_{\rm J}\sim{\cal O}({\cal L}/\lambda_J)$ and the weight-function ${\cal D}(\boldsymbol{x},t)$, whose detailed form can be found in \citet{circ_drag}, defines the region of influence which sound waves sent off by the perturber do not have time to leave. From the plane-symmetry of the problem, it is clear that ${\cal D}(x,y,z,t)={\cal D}(x,y,-z,t)$
and $\varepsilon_{\rm J}(x,y,z,t)=\varepsilon_{\rm J}(x,y,-z,t)$. Moreover, from the gradient of equation \eqref{eq:wave_circ} it also follows that
\eq
\partial_t{\cal D}=-\Omega\,\partial_\theta{\cal D}\;.
\eeq
If we are again concerned with the force exerted by the fluid, which includes dynamical friction effects, 
rather than with the accretion drag or the self-force
\footnote{Note that, differently from the case of straight-line motion, even the dissipative part
of the self-force is now non-zero, as can be seen from the fact that the perturber
loses energy and angular momentum through gravitational waves (\textit{cf.} the quadrupole formula).
Self-force calculations, as already mentioned, require different techniques \citep{poisson_rev} and can be performed separately.},
we can restrict our attention to the metric perturbations $\phi^{\rm F}=\psi^{\rm F}$
generated by the fluid, which are again given by equation \eqref{eq:poisson_eq}. Using again 
the Green's function of the Yukawa operator and evaluating at the position of the perturber
($r=R$, $\theta=\Omega t$, $z=0$) 
we easily get
\begin{equation}
\label{eq:grad_z_circ} \partial_z\phi^{\rm F}=\partial_z\psi^{\rm F}=\varepsilon_{\rm not\,DF}
\end{equation}
(from the plane symmetry of the function $\cal D$). For the azimuthal gradient, instead, we have
\begin{align}\label{eq:grad_theta_circ}
&\partial_\theta\psi^{\rm F}=\partial_\theta\phi^{\rm F} =-\frac{\partial_t\psi^{\rm F}}{\Omega}=-\frac{\partial_t\phi^{\rm F}}{\Omega}=\\
&(p+\rho)\int d^3{x}' \frac{\frac{\delta n}{n}\left(\boldsymbol{x}',t={\theta}/{\Omega}\right)(\boldsymbol{x}- \boldsymbol{x}')\cdot\partial_\theta \boldsymbol{x}}{\vert \boldsymbol{x}-\boldsymbol{x}'
\vert^{3}}\nonumber\\& \times(1+\varepsilon_{\rm Yukawa})+\varepsilon_{\rm not\,DF}=\nonumber\\
&\frac{4\pi(p+\rho)M\gamma(1+v^2)R}{v^2}\,I_\theta\times[1+{\cal O}({\cal L}/\lambda_J)]+\varepsilon_{\rm not\,DF}\;,\nonumber
\end{align}
where $I_\theta$ is given by
\begin{equation}
I_\theta\equiv-\frac{{\cal M}^2}{4\pi}\int d^3{\hat{x}}'\frac{{\cal D}\left(\boldsymbol{x}',t={\theta}/{\Omega}\right)\hat{r}'\sin(\theta'-\theta)}{[1+\hat{z}'^2+\hat{r}'^2-2\hat{r}'\cos(\theta-\theta')]^{3/2}}
\end{equation}
(a hat denotes quantities scaled by the radius of the orbit: $\hat{\boldsymbol x}'\equiv {\boldsymbol x}'/R$, $\hat{r}'=r'/R$, $\hat{z}'=z'/R$).
Similarly, for the radial gradient we obtain
\begin{align}
&\partial_r\psi^{\rm F}=\partial_r\phi^{\rm F} =\label{eq:grad_r_circ}\\
&(p+\rho)\int d^3{x}' \frac{\frac{\delta n}{n}\left(\boldsymbol{x}',t={\theta}/{\Omega}\right)(\boldsymbol{x}-\boldsymbol{x}')\cdot\partial_r \boldsymbol{x} }{\vert \boldsymbol{x}-\boldsymbol{x}'\vert^{3}}
\nonumber\\& \times(1+\varepsilon_{\rm Yukawa})+\varepsilon_{\rm not\,DF}=\nonumber\\
&\frac{4\pi(p+\rho)M\gamma(1+v^2)}{v^2}\,I_r\times[1+{\cal O}({\cal L}/\lambda_J)]+\varepsilon_{\rm not\,DF}\nonumber\;,
\end{align}
where $I_r$ is given by
\begin{equation}
I_r\equiv-\frac{{\cal M}^2}{4\pi}\int d^3{\hat{x}}'\frac{{\cal D}\left(\boldsymbol{x}',t={\theta}/{\Omega}\right)[\hat{r}'\cos(\theta-\theta')-1]}{[1+\hat{z}'^2+\hat{r}'^2-2\hat{r}'\cos(\theta-\theta')]^{3/2}}\;.
\end{equation}
Note that the integrals $I_\theta$ and $I_r$ have been calculated numerically in \citet{circ_drag}.
They are functions of the coordinate $\theta$ of the perturber, which is thought to vary in
an unbound range to count the number of revolutions, or equivalently they can be thought of as functions of time ($t=\theta/\Omega$). 
Fortunately, though, steady state values for these integrals are reached in times comparable 
to the sound crossing-time $R/c_s$ or within one orbital period: fits to the numerical results for these steady state values 
in the case in which $R\gg r_{\min}$ and ${\cal L}\gtrsim (20-100) R$ 
are given by \citep{circ_drag}
\begin{equation}\label{eq:IR}
  I_r = \left\{\begin{array}{l}
    {\cal M}^2\ 10^{\ 3.51{\cal M}-4.22},\quad   \mbox{for }{\cal M}<1.1\,,\\\\
    0.5\  \ln\big[9.33{\cal M}^2({\cal M}^2-0.95)\big],\\ \mbox{for }1.1\leq{\cal M}<4.4\,,\\\\
    0.3\ {\cal M}^2, \quad  \mbox{for }{\cal M}\geq4.4\,,
  \end{array}\right.
\end{equation}
and
\begin{equation}\label{eq:Iphi}
  I_\theta = \left\{\begin{array}{l}
    0.7706\ln\left(\frac{1+{\cal M}}{1.0004-0.9185{\cal M}}\right)-1.4703{\cal M}, \\\mbox{for }{\cal M}<1.0\,,\\\\
     \ln[330 (R/r_{\rm min}) ({\cal M}-0.71)^{5.72}{\cal M}^{-9.58} ],\\\mbox{for }1.0\leq{\cal M}<4.4\,, \\\\
     \ln[(R/r_{\rm min})/(0.11{\cal M}+1.65)],\\ \mbox{for }{\cal M}\geq4.4\,.
  \end{array}\right.
\end{equation}
These fits are accurate within 4\% for ${\cal M}<4.4$ and within 16\% for ${\cal M}>4.4$.

Using equations \eqref{eq:grad_z_circ},  \eqref{eq:grad_theta_circ} and \eqref{eq:grad_r_circ} in equation \eqref{eq:acceleration} 
and transforming to cylindrical coordinates, for the acceleration produced by the gravitational interaction with the fluid we easily get
\begin{align}
\label{eq:a_t}\left(\tilde{a}^t_{\rm pert}\right)_{\rm F}=\,&-\frac{4\pi(p+\rho)M\gamma^3(1+v^2)^2}{v}\,I_\theta\\&\times[1+{\cal O}({\cal L}/\lambda_J)+{\cal O} (M/r_{\min})]+\varepsilon_{\rm not\,DF}\,,\nonumber\\
\left(\tilde{a}^\theta_{\rm pert}\right)_{\rm F}=\,&-\frac{4\pi(p+\rho)M\gamma^3(1+v^2)^2}{R\,v^2}\,I_\theta\\&\times[1+{\cal O}({\cal L}/\lambda_J)+{\cal O} (M/r_{\min})]+\varepsilon_{\rm not\,DF}\,,\nonumber\\
\left(\tilde{a}^r_{\rm pert}\right)_{\rm F}=\,&-\frac{4\pi(p+\rho)M\gamma^3(1+v^2)^2}{v^2}\,I_r\\&\times[1+{\cal O}({\cal L}/\lambda_J)+{\cal O} (M/r_{\min})]+\varepsilon_{\rm not\,DF}\,,\nonumber\\
\label{eq:a_z}\left(\tilde{a}^z_{\rm pert}\right)_{\rm F}=\,\,&\varepsilon_{\rm not\,DF}\,.
\end{align}
[The error ${\cal O} (M/r_{\min})$ comes about because the equations that we have solved 
are linearized and are therefore subject to
an ``intrinsic'' error ${\cal O} (\varepsilon_1,\varepsilon_2)$.]

Finally, in order to compute the change of 3-momentum due to the
gravitational interaction with the fluid in the rest frame of the perturber, it is sufficient to project the 4-force $M(\tilde{a}^\mu_{\rm pert})_{\rm F}$
onto a tetrad comoving with the perturber, \textit{i.e.,} $e_{(t)}=u_{\rm pert}^\mu\partial/\partial x^\mu=\gamma(\partial/\partial t+\Omega \partial/\partial \theta)$,
$e_{(\theta)}=\gamma(v\partial/\partial t+1/r\, \partial/\partial \theta)$, $e_{(r)}=\partial/\partial r$ and $e_{(z)}=\partial/\partial z$.
Using equations \eqref{eq:a_t}--\eqref{eq:a_z} one then easily gets
\begin{align}
  &\left(\frac{d \tilde{p}_{\rm pert}^{(\theta)}}{d\tau}\right)_{\rm F}=-\frac{4 \pi (p+\rho) M^2\gamma^2 (1+v^2)^2}{v^2}\, I_\theta\nonumber\\
&\times  [1+{\cal O} ({\cal L}/\lambda_J)+{\cal O} (M/r_{\min})]+\varepsilon_{\rm not\,DF}\label{eq:tangential_drag}
\,,\\
 & \left(\frac{d \tilde{p}_{\rm pert}^{(r)}}{d\tau}\right)_{\rm F}=-\frac{4 \pi (p+\rho) M^2\gamma^3 (1+v^2)^2}{v^2}\, I_r\nonumber\\
&\times [1+{\cal O} ({\cal L}/\lambda_J)+{\cal O} (M/r_{\min})]+\varepsilon_{\rm not\,DF}\label{eq:radial_drag}
\,,\\
 & \left(\frac{d \tilde{p}_{\rm pert}^{(z)}}{d\tau}\right)_{\rm F}=\,\varepsilon_{\rm not\,DF}\,.
\end{align}
As in the case of straight-line motion, the relative errors ${\cal O} ({\cal L}/\lambda_J)$ 
and  ${\cal O} (M/r_{\min})$ are negligible, because the fluid is not self-gravitating
and because the effective cutoff radius $r_{\min}$ is large compared with $M$, 
whereas $\varepsilon_{\rm not\,DF}$ in general is not negligible, 
although it is exactly zero if the medium is spherically symmetric around the perturber.
The relativistic correction factors  $\gamma^2(1+v^2)^2$ and $\gamma^3(1+v^2)^2$ appearing in equations \eqref{eq:tangential_drag} 
and \eqref{eq:radial_drag} are plotted as functions of the velocity $v$ in Fig. \ref{fig:corrections}.
\section{\label{sec:conclusions}Conclusions}

We have studied the drag experienced by a massive body  
because of the gravitational interaction with its own
gravitationally-induced wake,
when it is moving along 
a straight-line or a circular orbit 
 at relativistic speed $v$ relative to
 a non self-gravitating collisional fluid in a flat 
or weakly curved background spacetime.
Thanks to a suitable choice of gauge, we could exploit the Newtonian 
analysis of \citet{ostriker} and of \citet{circ_drag} to simplify our calculations. 
We find that their results remain valid also in the 
relativistic case, provided that the rest-mass density is replaced by $p+\rho$ ($p$ and $\rho$ being the 
pressure and energy density of the fluid) and a relativistic multiplicative factor is included. 
This factor turns out to be $\gamma^2[1+(v/c)^2]^2$ 
in the straight-line motion case and for the tangential component of the drag in the circular motion case, and 
$\gamma^3[1+(v/c)^2]^2$ for the radial component of the drag in the circular motion case. 

Although our analysis strictly applies only to a fluid in a flat spacetime (in the case
of straight-line motion) or a weakly curved one (in the case of circular motion), 
we have argued that our results are suitable at least for a preliminary study 
of the effects of an accretion disc on EMRIs.
Although our results are not expected to change the standard conclusion that 
the gas accreting onto the central SMBH does not significantly affect EMRIs in the case of ``normal'' Galactic Nuclei \citep{narayan},
they could play a role, under certain circumstances, in the case of higher density environments like Active
Galactic Nuclei (quasars, Seyfert Galaxies, etc.). An investigation of this scenario, in which the accretion
is modelled by a thick torus, will be presented in a subsequent paper \citep{nextpaper}.

\section*{Acknowledgments}

I am grateful to L.\ Rezzolla for suggesting this problem
and making stimulating comments about this work, and to 
J.\ C.\ Miller for carefully reading this manuscript and
giving helpful advice on it.
I would also like to thank M.\ Cook (as well as J.\ C.\ Miller)
for carefully checking the language of this manuscript, helping
to improve it far above the level of my own skills.

\label{lastpage}


\begin{thebibliography}{99}

\bibitem[\protect\citeauthoryear{Amaro-Seoane et 
al.}{2007}]{emris} 
Amaro-Seoane P., Gair J.~R., Freitag M., 
Miller M.~C., Mandel I., Cutler C.~J., Babak S., 2007, CQGra, 24, 113 

\bibitem[\protect\citeauthoryear{Babak et al.}{2006}]{kludge} 
Babak S., Fang H., Gair J.~R., Glampedakis K., Hughes S.~A., 2007, PhRvD, 
75, 024005 

\bibitem[\protect\citeauthoryear{Barack \& Ori}{2001}]{gauge} 
Barack L., Ori A., 2001, PhRvD, 64, 124003 

\bibitem[\protect\citeauthoryear{Barausse}{2007}]{nextpaper} Barausse E., 2007, in preparation 

\bibitem[\protect\citeauthoryear{Bardeen}{1980}]{bardeen} 
Bardeen J.~M., 1980, PhRvD, 22, 1882 

\bibitem[\protect\citeauthoryear{Binney \& Tremaine}{1987}]{BT} Binney J., Tremaine S., 1987, Galactic dynamics.
Princeton University Press, Princeton, NJ

\bibitem[\protect\citeauthoryear{Blandford \& Begelman}{1999}]{ADIOS} Blandford R.~D., Begelman M.~C., 
1999, MNRAS, 303, L1 

\bibitem[\protect\citeauthoryear{Bondi \& 
Hoyle}{1944}]{bondi1} Bondi H., Hoyle F., 1944, MNRAS, 104, 
273 

\bibitem[\protect\citeauthoryear{Bondi}{1952}]{bondi2} Bondi 
H., 1952, MNRAS, 112, 195 

\bibitem[\protect\citeauthoryear{Chakrabarti}{1993}]{chakra1} 
Chakrabarti S.~K., 1993, ApJ, 411, 610

\bibitem[\protect\citeauthoryear{Chakrabarti}{1996}]{chakra2} 
Chakrabarti S.~K., 1996, PhRvD, 53, 2901 

\bibitem[\protect\citeauthoryear{Chandrasekhar}{1943}]{chandra} 
Chandrasekhar S., 1943, ApJ, 97, 255 

\bibitem[\protect\citeauthoryear{Flanagan \& Hughes}{2005}]{scott} Flanagan {\'E}.~{\'E}., Hughes S.~A., 
2005, NJPh, 7, 204 

\bibitem[\protect\citeauthoryear{Karas \& {\v 
S}ubr}{2001}]{subr4} Karas V., {\v S}ubr L., 2001, A\&A, 376, 
686 

\bibitem[\protect\citeauthoryear{Kim \& Kim}{2007}]{circ_drag} 
Kim H., Kim W.-T., 2007, ApJ, 665, 432 

\bibitem[\protect\citeauthoryear{King, Pringle, \& 
Livio}{2007}]{alpha_value} King A.~R., Pringle J.~E., Livio M., 
2007, MNRAS, 376, 1740 

\bibitem[\protect\citeauthoryear{Kodama \& 
Sasaki}{1984}]{kodama_sasaki} Kodama H., Sasaki M., 1984, PThPS, 78, 1 

\bibitem[\protect\citeauthoryear{Lee}{1969}]{PNdf} Lee 
E.~P., 1969, ApJ, 155, 687 


\bibitem[\protect\citeauthoryear{Ma \& 
Bertschinger}{1995}]{ma_bertschinger} Ma C.-P., Bertschinger E., 1995, 
ApJ, 455, 7 

\bibitem[\protect\citeauthoryear{Mino}{2003}]{mino} Mino Y., 
2003, PhRvD, 67, 084027 


\bibitem[\protect\citeauthoryear{Misner, Thorne, \& 
Wheeler}{1973}]{gravitation} Misner C.~W., Thorne K.~S., Wheeler 
J.~A., 1973, Gravitation, W.H.~Freeman and Co., San Francisco, USA

\bibitem[\protect\citeauthoryear{Mukhanov, Feldman, \& 
Brandenberger}{1992}]{mukhanov} Mukhanov V.~F., Feldman H.~A., 
Brandenberger R.~H., 1992, PhR, 215, 203 

\bibitem[\protect\citeauthoryear{Narayan \& Yi}{1994}]{ADAF} 
Narayan R., Yi I., 1994, ApJ, 428, L13 

\bibitem[\protect\citeauthoryear{Narayan}{2000}]{narayan} 
Narayan R., 2000, ApJ, 536, 663 

\bibitem[\protect\citeauthoryear{Ostriker}{1999}]{ostriker} 
Ostriker E.~C., 1999, ApJ, 513, 252 

\bibitem[\protect\citeauthoryear{Pati \& Will}{2000}]{will_pati} 
Pati M.~E., Will C.~M., 2000, PhRvD, 62, 124015 

\bibitem[\protect\citeauthoryear{Petrich et 
al.}{1989}]{petrich} Petrich L.~I., Shapiro S.~L., Stark R.~F., 
Teukolsky S.~A., 1989, ApJ, 336, 313 

\bibitem[\protect\citeauthoryear{Poisson}{2004}]{poisson_rev} 
Poisson E., 2004, LRR, 7, 6 


\bibitem[\protect\citeauthoryear{Rephaeli \& 
Salpeter}{1980}]{rephaeli_salpeter} Rephaeli Y., Salpeter E.~E., 1980, 
ApJ, 240, 20 

\bibitem[\protect\citeauthoryear{Ruderman \& 
Spiegel}{1971}]{wakes} Ruderman M.~A., Spiegel E.~A., 1971, 
ApJ, 165, 1 

\bibitem[\protect\citeauthoryear{S{\'a}nchez-Salcedo \& 
Brandenburg}{1999}]{simulations} S{\'a}nchez-Salcedo F.~J., 
Brandenburg A., 1999, ApJ, 522, L35 

\bibitem[\protect\citeauthoryear{Shakura \& 
Sunyaev}{1973}]{thin_disk} Shakura N.~I., Sunyaev R.~A., 1973, 
A\&A, 24, 337 

\bibitem[\protect\citeauthoryear{Syer}{1994}]{syer} Syer D., 
1994, MNRAS, 270, 205 


\bibitem[\protect\citeauthoryear{{\v S}ubr \& 
Karas}{1999}]{subr3} {\v S}ubr L., Karas V., 1999, A\&A, 352, 452

\bibitem[\protect\citeauthoryear{Vokrouhlicky \& 
Karas}{1993}]{subr1} Vokrouhlicky D., Karas V., 1993, MNRAS, 265, 365

\bibitem[\protect\citeauthoryear{Vokrouhlicky \& 
Karas}{1998}]{subr2} Vokrouhlicky D., Karas V., 1998, MNRAS, 298, 53

\bibitem[\protect\citeauthoryear{Walker \& 
Will}{1980}]{will_walker} Walker M., Will C.~M., 1980, ApJ, 242, 
L129 




\end{thebibliography}
\end{document}